\documentstyle[aps,pra,epsfig,twocolumn]{revtex}
\begin{document}

\wideabs{
\title{Vortex bending and tightly packed vortex lattices in Bose-Einstein condensates}

\author{J.~J. Garc{\'\i}a--Ripoll \cite{emiliojjgr} \& V.~M. P\'erez-Garc{\'\i}a
\cite{emiliovmpg}}

\address{Departamento de Matem\'aticas, E. T. S.
  I. Industriales, Universidad de Castilla-La Mancha,\\ Av. Camilo Jos\'e Cela 3, 13071 Ciudad
  Real, Spain }

\maketitle

\draft

\begin{abstract}
  We study in detail the structure of the ground state of elongated
  rotating Bose--Einstein condensates. Such ground state is composed
  of one or more vortex lines which bend even in completely symmetric
  setups.  This symmetry breaking allows the condensate to smoothly
  adapt to rotation and to produce tightly packed arrays of vortex
  lines. The dependence of vortex bending with respect to relevant
  parameters is studied.
\end{abstract}

\pacs{PACS number(s): 03.75.Fi, 05.30.Jp, 67.57.De, 67.57.Fg}

}

\section{Introduction}

A stirred coffee climbing up the cup walls, a whirl of water spiraling down a
drain, or a tornado walking over the earth, are natural phenomena involving the
rotation of a fluid, let it be coffee, water or dusty air. In all of them the
fluid is subject to a balance between an external force and the centrifugal
force, and this balance is responsible for a depression of the density along
the axis of rotation. These structures have no trivial symmetries as is evident
in the case of tornadoes, which have continuously changing bent shapes.  None
of these structures have long lives because viscosity, imperfections and other
dissipative mechanisms play such an important role that rotation cannot be self
sustained.

This behavior contrasts with that of the so called superfluids, which represent
a state of matter of negligible viscosity or {\em dry fluid} \cite{neumann}.
The lack of viscosity allows a superfluid to host rotation for long times with
little or no external intervention. Besides, when a superfluid rotates it
follows a special type of flow which is irrotational, $\nabla \times {\bf v} =
0$, except for a finite number of extended singularities, which are called {\em
  vortex lines}.  These vortex lines, first predicted for $^4$He condensates,
are the superfluid equivalent of robust whirls and tornadoes.  However, due to
centrifugal forces, the density of the superfluid \cite{nota} becomes zero
along the vortex line something which is difficult to achieve in ordinary
liquids.

Until recently only two examples of true superfluids were known: the superfluid
phases of $^4$He and $^3$He. In both cases, the strength of the many--body
interactions obscure the signatures of superfluidity and make their study
difficult. The search for superfluid--type, weakly interacting systems has led
to the experimental realizations of Bose--Einstein condensates (BEC) using
dilute gases of alkali atoms \cite{BEC}. These condensates are ruled at low
temperature by a mesoscopic quantum wavefunction, $\psi({\bf r}) =
\sqrt{\rho({\bf r})}e^{i\theta({\bf r})}$, whose phase determines a velocity
field, ${\bf v}=\hbar\nabla \theta({\bf r})/m$, that may host vortices along
the lines where $\theta$ is not well defined. The existence of vortices in BEC
has been directly confirmed in two sets of experiments. The first one is based
on the manipulation of the internal degrees of freedom of an almost spherical
condensate \cite{JILA}. In the second one, vortices are produced by
``mechanical means'' in rotating elongated traps with transverse asymmetries
\cite{ENS1,ENS2}, a geometry that leads to intriguing effects on the vortex
nucleation process.

\begin{figure}
  \begin{center}
     \epsfig{file=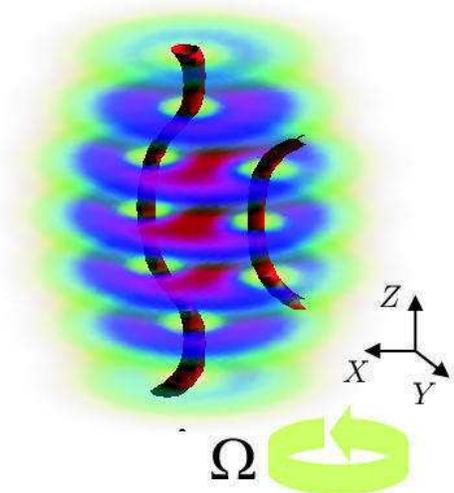,width=0.9\linewidth}
  \end{center}
  \caption{\label{fig-0}
    Ground state of a condensate in an asymmetric trap ($\varepsilon=0.03$) as
    the one from current experiments for an angular speed of $\Omega =
    0.55\omega_\perp$. Shownare the vortex cores and several transverse
    sections of the condensate.  The gray scale indicates the modulus of the
    wavefunction. We also indicate the axis of rotation of the trap.}
\end{figure}

In Ref. \cite{Butts} we find several commonly accepted predictions about
vortices in rotating traps. Namely, (i) a rotating trap leads to the production
of one or more vortices of unit topological charge $m$, being $m =
(2\pi)^{-1}\oint \nabla\theta$.  (ii) Vortices appear in finite numbers, with
straight shapes, forming lattices with p--fold symmetries.  (iii) Certain
critical rotation speeds $\Omega_p<\Omega_{p+1}$ should be surpassed before
each new vortex is nucleated. (iv) Asymmetric states without rotational
symmetry of any kind are found to be energetically unstable, which means that
there are perturbations with no energetical cost that can destroy such states.
However, Ref.  \cite{Butts} is based on a variational ansatz for the weak
coupling limit, which lacks longitudinal degrees of freedom and implicitly
induces the $p$--fold symmetry of the vortex lattice.  Recent works abandoned
these constraints but either focused on infinitely long condensates or studied
almost spherically symmetric traps \cite{others}.

In this paper we study an elongated three--dimensional Bose--Einstein
condensate subject to rotation. Our main result is that stationary vortex lines
are bent even in completely symmetric setups and that the bending depends on
the type of trap on the number of atoms of the experiment. The fact that vortex
lines may bend was mentioned in Ref. \cite{Fetter} where longitudinal
excitation modes of a vortex line induce vortex bending. However the physics in
that paper differs drastically from our work, in which we consider {\em the
  ground state} and the generation of {\em stable and stationary} structures.
Second, in a complementary paper \cite{Ours} we have presented some evidences
in favor of that bending and how it could be related to the experimental
results of the ENS \cite{ENS1}. In this work we extend that work to the
analysis of arrays of bent vortices and its properties as well as the
dependence of the bending of the vortices on the geometry of the trap.

Our work consists on two parts. In Section \ref{sec:tools} we provide the
mathematical foundations for the search of the rotating condensate most
favorable configurations. To do so we first introduce the mean-field model used
to describe the condensate together with some basic definitions and a new
energy functional, whose minima are the possible ground states of the
condensate. In this section we also develop an efficient minimization method
for this functional.

All these tools are then applied in Sect. \ref{sec:results} to different
configurations. We start with setups from current experiments \cite{ENS1} and
show that in elongated traps with many bosons the vortex lines are bent and
form Abrikosov lattices which are regular only in the core of the gas cloud and
deform close to the boundaries. Next we study the dependence of the vortex
bending with respect to the free parameters in our theory. Finally in Sect.
\ref{sec:final} we offer our conclusions together with some open questions.

\section{Mathematical tools}
\label{sec:tools}

\subsection{The mean-field model}
\label{sec:mean-field}

For current experiments it is an accurate approximation to use the zero
temperature mean field theory of the condensate, in which the atomic cloud is
described by a single wavefunction $\psi({\bf r},t)$ ruled by a
Gross-Pitaevskii equation (GPE).  In the case of rotating systems it is useful
to consider the problem on the mobile reference frame that moves with the trap
\cite{asym}, in which the equation reads
\begin{equation}
\label{GPE-rot} i\frac{\partial \psi}{\partial t}  =\left[
-\frac{1}{2}\triangle + V_0({\bf r}) + g\left| \psi \right| ^{2}
-\Omega L_{z}\right] \psi.
\end{equation}
Here $L_{z}=i\left(x\partial_y-y\partial_x\right)$ is the hermitian operator
that represents the angular momentum along the z-axis and the effective
trapping potential in given by
\begin{equation}
V_0({\bf r}) = \frac{1}{2}\omega_\perp^2\left[(1-\varepsilon)x^2+
(1+\varepsilon)y^2]+\frac{1}{2}\omega_z^2z^2\right].
\end{equation}
In Eq. (\ref{GPE-rot}) we have applied a convenient adimensionalization which
uses the harmonic oscillator length, $a_\perp=\sqrt{\hbar/m_{Rb}\omega_\perp}$
and period, $\tau =\omega_\perp^{-1}.$ With these units the nonlinear parameter
becomes $g=4\pi a_{\text{S}}/a_\perp$.

\subsection{Variational formulation}
\label{sec:energy}

There are several conserved quantities associated to Eq.  (\ref{GPE-rot}). The
first one is the norm of the wavefunction, $N[\psi]=\int \left|\psi\right|^2
d{\bf r}$, which is related to the number of bosons in the condensate. The
second conserved equation is the energy of the gaseous condensate
\begin{eqnarray}
E[\psi ] & = & \int \bar{\psi}\left[-\frac{1}{2}\triangle
+V_{0}\left({\bf r}\right)
+\frac{g}{2}\left|\psi\right|^2-\Omega L_{z}\right]\psi d{\bf r} \nonumber \\
& = & E_0(\psi) - \Omega  L_z(\psi). \label{energy}
\end{eqnarray}

In current experiments with stirred condensates \cite{ENS1,ENS2,abo-shaer}, the
gaseous cloud reaches certain long lived configurations with one ore more
vortices. In this work we are not interested on the precise dynamics of the
nucleation process, but look for the final stationary configurations. Due to
thermodynamical considerations we expect the condensate to achieve the least
energy for given experimental parameters, $\{N,g,\Omega\}$. Such configuration
is called a \emph{ground state}.

Within the framework of our mean-field model, a stationary state is represented
by a wavefunction with the form
\begin{equation}
\label{static}
\psi_{\mu}({\bf r},t) =e^{-i\mu t}\phi({\bf r}).
\end{equation}
The first way to find such solutions is to introduce (\ref{static}) in our
mean--field model (\ref{GPE-rot}) and directly solve the resulting partial
differential equation. This method has several disadvantages: (i) It is
difficult due to the many degrees of freedom that it involves, (ii) this new
equation is satisfied not only by the ground state but also by excited states,
and (iii) there is no mathematical guarantee yet for the existence of solutions
of this problem.

The second way to characterize a stationary is solution is by studying the
energy functional and use the fact that each stationary configuration is a
critical point of the energy (\ref{energy}) for given $\{N,g,\Omega\}$,
\begin{equation}
  \left. \frac{\partial E}{\partial \psi }\right|_N[\psi_\mu]=0.
\end{equation}
Since we are indeed looking for the ground state, one should use a minimization
procedure able to find the minima of the energy functional for given
parameters.

\subsection{Reshaping the energy functional}
\label{sec:free-energy}

As we stated above, our objective is to find the solutions $\psi_\mu$ which are
the minima of the energy subject to the restriction $\int |\psi|^2\equiv N$,
for a fixed {\em angular speed of the trap} and for a given interaction. The
existence of at least one minima or {\em ground state} for this variational
problem has been proved elsewhere \cite{Sobolev}, they are guaranteed to be
stable and represent the energetically most favorable configurations for given
$N$ and $\Omega$.

Such a problem is called a constrained optimization and we will refer to the
solution, $\psi_\mu$, as a constrained minimum. The existence of constraints in
an optimization problem poses serious difficulties to traditional descent
methods, since it is difficult to design a efficient minimization algorithm
which takes care of the constraints at each step \cite{nota2}. To take into
account this restriction one should use Lagrange's multipliers, i.e. just
adding a fraction of the constraint $\omega(\mu,N)$ (also called a
``penalizer'') to the original functional
\begin{equation}
  F[\psi] = E[\psi] + \omega(\mu,N).
\end{equation}
It is not difficult to show that $F[\psi]$ and $E[\psi]$ have the same
stationary states, and that any absolute or relative minimum of $F[\psi]$ is
also a constrained minimum of $E[\psi]$.

One expects that a wise choice of the penalizer will establish a one-to-one
correspondence between the values of the chemical potential, $\mu$, the
absolute minima of $F[\psi]$ and the ground states of our condensate. The
advantage of $F[\psi]$ over $E[\psi]$ is that in the new functional there is no
need to care for constraints: the value of $N$ for the ground state is
determined by the chemical potential, $\mu$, which is fixed throughout the
minimization process.

The traditional choice for a Lagrange multiplier leads to the definition of the
free energy
\begin{equation}
  {\cal F}[\psi] = E[\psi] - \mu N[\psi].
\end{equation}
However, this functional is not bounded below and thus $\psi_\mu$ is at most a
local minimum of ${\cal F}$. Consequently, this functional cannot be used to
characterize the state of the condensate. This is an interesting result, since
one could be tempted to think that the traditional definitions of the
thermodynamical potentials are suitable to characterize all physical systems.
We must remember, however, that our condensate is being described by a
mean-field model, a simplification of a more complex model, and as such what
works for the full problem needs not work for the simpler one.

In our search for suitable functionals we have found a simple one which we call
the {\em nonlinear free energy}
\begin{equation}
\label{new-free-energy}
  F[\psi]=E[\psi]+\frac{1}{2}\left(N[\psi]-\lambda \right)^2.
\end{equation}
First and most important, it can be proven that $F[\psi]$ has at least one
finite norm absolute minimum for each value of $\lambda$ \cite{Sobolev}, and
that each of those minima corresponds to a constrained minima of the energy,
$E[\psi]$. And second, there exists a simple and invertible relation between
the usual thermodynamic variables ---the chemical potential and the number of
particles, $(\mu,N)$---, and our new variables ---Lagrange's constant and the
number of particles, $(\lambda,N)$---,
\begin{equation}
  \mu =N[\psi_\mu]-\lambda.
\end{equation}

An important feature of our new functional (\ref{new-free-energy}) is that it
highly nonlinear with respect to $\psi$. This poses no new difficulty, since
our original equations (\ref{GPE-rot}) were already nonlinear. Indeed
Eq. (\ref{new-free-energy}) has proven to be the most natural choice for
many other problems, such as the propagation of incoherently coupled laser
beams through saturable media \cite{dipoles}.

\subsection{Optimal methods for minimization}
\label{sec:sobolev}

To minimize the nonlinear free energy (\ref{new-free-energy}) we follow Ref.
\cite{Neu97}. First we choose the right function space, which in our case
is the Sobolev space $H^1(R^3) \equiv \{\psi/\psi,\nabla \psi\in L^2\}$ of
functions which admit at least one spatial derivative. This space is equipped
with a scalar product
\begin{equation}
  \langle \psi,\phi \rangle  \equiv
  \int \left[ \bar{\psi}({\bf r})\phi({\bf r})+\nabla
  \bar{\psi}({\bf r})\cdot \nabla \phi({\bf r})\right] d^nr.
\end{equation}
and a norm $\Vert\psi\Vert= \langle\psi,\psi\rangle$.  To obtain an explicit
expression for the gradient of the functional in $H^1$ we perform a first order
expansion of $F[\psi]$ around a trial state $\psi$
\begin{eqnarray}
F\left[\psi+\epsilon \delta \right]&=& F[\psi] +2 \epsilon
\mathrm{Re}\int \left(\delta^*\frac{\partial E}{\partial
\bar{\psi}} +
  \nabla \delta^* \frac{\partial E}{\partial \nabla \psi}\right)
\nonumber\\
&+& O(\epsilon^2).
\end{eqnarray}
We have to turn this expression into something which looks like Frechet's
definition of a derivative. This means that we have to find some $\phi$ such
that
\begin{equation}
  \mathrm{Re}\int \left[ \bar{\delta }\frac{\partial E}{\partial \bar{\psi}}+\nabla
    \bar{\delta }\frac{\partial E}{\partial (\nabla \bar{\psi})}\right]
  =\mathrm{Re}\int \left[ \bar{\delta }\phi +\nabla \bar{\delta }\nabla \phi \right].
\end{equation}
If we integrate by parts and impose that this equality be satisfied for all
perturbations, $\delta$, the problem has a formal solution which is given by a
Lagrange equation
\begin{equation}
  \left( 1-\triangle \right) \phi =\frac{\partial E}{\partial \bar{\psi}}
  -\nabla\frac{\partial E}{\partial (\nabla \bar{\psi})}.
\end{equation}
In consequence, our formal expression for the Sobolev gradient of $F[\psi]$
finally reads $\nabla_SF\equiv \left( 1-\triangle \right) ^{-1}\nabla F,$ where
$\nabla_SF$ stands for the Sobolev gradient, $\nabla F$ is the ordinary one,
and $(1-\triangle)^{-1}$ represents the inverse of a linear and strictly
positive definite operator.

We may now use the Sobolev gradient of our functional as the direction of
descent for a minimization procedure
\begin{eqnarray}
\frac{\partial \nu }{\partial \tau}({\bf r},\tau) = \left(
1-\triangle\right)^{-1}\left[ -\frac{1}{2}\triangle +V+ g|\nu
|^2-\Omega L_z\right] \nu.\label{bec-free-energy-sobolev}
\end{eqnarray}
The preceding equation converges to some stationary state $\phi({\bf r}) =
\lim_{\tau \rightarrow \infty} \nu({\bf r},\tau)$. To grant convergence to the
true ground state one needs to improve the descent method, using for instance
the nonlinear conjugate gradient method instead of
(\ref{bec-free-energy-sobolev}), and introducing some type of relaxation which
helps avoid saddle node points. In any case, the minimization process must be
performed on suitable space of functions that we built using a discrete Fourier
basis with $64\times{64}\times{128}$ modes.

\begin{figure}
  \begin{center}
    \epsfig{file=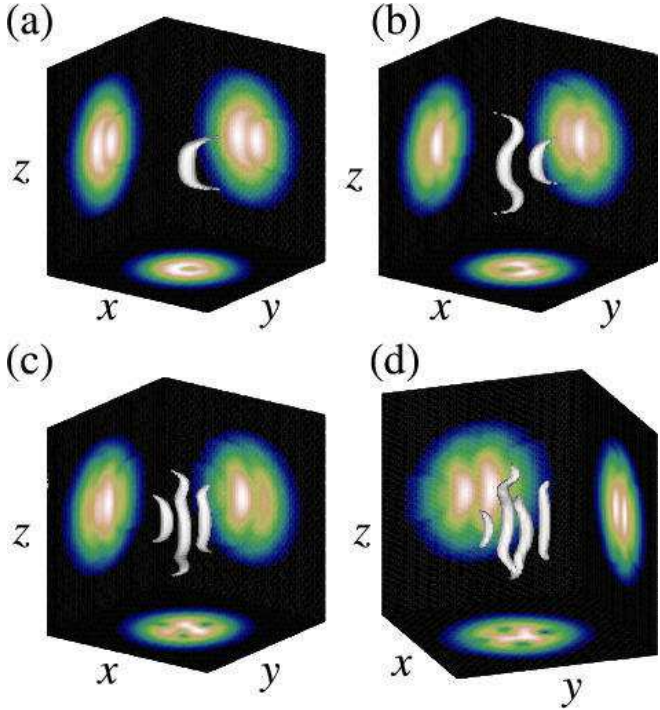,width=\linewidth}
  \end{center}
  \caption{\label{fig-1}
    Three--dimensional surface plots of one to four vortex lines for
    $\varepsilon=0.03$ and $\Omega/\omega_\perp=0.41,0.55,0.6,0.65$.  The walls
    of the boxes reflect different integrated views of the gas cloud (i.e. as
    they would look if imaged with high resolution), and within each box we
    show the lines of lowest density within the condensate. Each plot is
    represents a centered box $(x,y,z)\in[-6,6]\times[-6,6]\times[-60,60]$,
    with sizes in adimensional units. Parameters are close to experimental
    values: $Ng=9000$, $\varepsilon=0.03$, $\omega_\perp/\omega_z=219/11.7$.}
\end{figure}

\section{Results}
\label{sec:results}

\subsection{Bent vortex lattices in current experiments}
\label{sec:bending}

We have applied the numerical methods outlined above to different setups. The
family of numerical experiments in this subsection resembles the experiments of
Madison et al. \cite{ENS1,ENS2}. For these experiments with $^{87}$Rb
condensates, the bosonic interaction is ruled by the scattering length
$a_{\text{S}}\simeq 5.5\,\textrm{nm}$. For several pictures in this paper, the
number of atoms was chosen to match, $Ng = 9000$, which corresponds to a few
times $10^5$ Rb bosons but our results remain qualitatively valid for an ample
range of $gN$ values as will be shown below.

The geometry of the trap is a very elongated one, $\omega_\perp > 18.7
\omega_z$, and for the small transverse deformation of the trap we have tried
$\varepsilon = 0.0$ as well as $\varepsilon = 0.03$, the latter being the
closest one to the actual experiment \cite{ENS1}. This means that we have
studied both \emph{axially symmetric setups and completely asymmetric setups},
and in both cases our study has provided essentially the same results. This is
due to the fact that an intense nonlinear interaction $(Ng \sim 10^4)$
effectively prolongs the existence of all solutions from the symmetric setup to
the asymmetric setup \cite{gabriel}.

The first result is that an increase of the angular speed causes the minimum of
the energy to move from a nodeless {\em ground state} into states with one,
two, three and more vortices.  Consequently, as it was already predicted
\cite{Butts}, there exists a cascade of increasing angular frequencies,
$\{\Omega_1 < \Omega_2 < \ldots\}$ for the nucleation of one, two and more
vortices ---the larger the rotation frequency, the more vortices, and for our
setup, $\Omega_1 \simeq 0.4(0)$, $\Omega_2\simeq 0.5(3)$, $\Omega_3 \simeq
0.5(8)$, and $\Omega_4\simeq0.6(4)$.  Figure \ref{fig-0} shows the structure of
a ground state hosting two vortices at $\Omega = 0.55$.

The second and most important result, which is already evident in Fig.
\ref{fig-0} is that vortices are nucleated with a {\em stable and stationary}
bent shape even in axially symmetric traps. In Fig. \ref{fig-1} we show
three--dimensional pictures of a condensate with up to four bent vortex lines.

These contorted shapes are {\em absolutely stable} configurations which lack
rotational symmetry of any kind, even discrete, something which was thought to
be forbidden according to Ref.  \cite{Butts}. In that case asymmetric states
were found only for discrete values of the angular speed, right on the
transition between different number of vortices, and always exhibiting
energetic instabilities. Furthermore, due to the bending, while the trunk of
the condensate rotates, the caps remain almost still.  This feature allows the
gas to accommodate a fractional value of the angular momentum which is between
0 and 1, and which never reaches 1, as it was already conjectured in
\cite{Ours}.  As the speed of the trap increases, the angular momentum grows
continuously by means of pulling the vortex line straighter and not only with
discontinuous jumps.

\subsection{Regularity of the vortex lattice}
\label{sec:cause}

From old studies with liquid helium and more recent works with condensates of
simpler geometry, it has been expected that vortices should form regular
triangular lattices. On the other hand, our finding above seems to suggest that
the vortex aggregate rather adopts an irregular shape. As we will show below,
the condensate actually develops a medium range order, which dilutes in the
vicinity of the cloud boundary.

\begin{figure}
  \begin{center}
    \epsfig{file=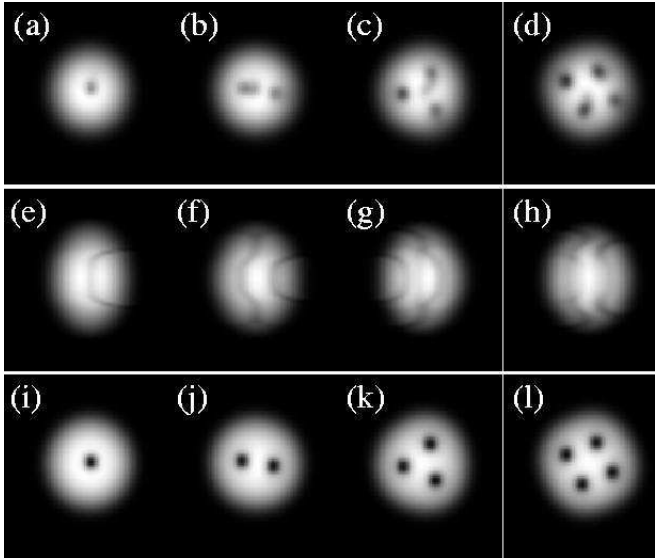,width=\linewidth}
  \end{center}
  \caption{\label{fig-2}
    Condensate shapes for the same states as in Fig. \ref{fig-1}:
    $\Omega/\omega_\perp=0.41$ (a,e,i) ,$\Omega/\omega_\perp=0.55$ (b,f,j),
    $\Omega/\omega_\perp=0.6$ (c,g,k) and $\Omega/\omega_\perp=0.65$ (d,h,l).
    Shownare up (a-d) and side views (e-h) and (i-l) two--dimensional cuts of
    the cloud at half its height.}
\end{figure}

In Fig. \ref{fig-2} we use the solutions of our previous section to simulate
the pictures that should be seen in experiments. The first row of
two--dimensional plots [Fig. \ref{fig-2}(a-d)] are upper views of the
condensate, and represent the column density of the bosonic cloud along the
transverse directions. These pictures bear an extreme resemblance with the
experimental photographs of Refs. \cite{ENS1,ENS2}, showing blurred clouds
where the vortex cores seem partially filled.

In Fig. \ref{fig-2}(i-l) we show the density of the condensate as seen at half
the height of the condensate, i.e.  $|\psi{(x,y,z=0)}|^2$. At the moment of
revising this paper an experiment has been made which observes this type of
transverse cuts of a stirred sodium condensate \cite{abo-shaer}. In that
experiment a thin slice of bosons is promoted to a different internal state and
then imaged separately from the rest of the cloud. The resulting pictures are
like those in Fig.  \ref{fig-2}(i-l), where the regularity of the Abrikosov
vortex lattice is made evident. The blurring has disappeared and the holes
arrange along the expected triangular lattice. In the above mentioned paper the
bending of the vortex lines is also reported.

As further confirmation we suggest that using the setup of Ref.  \cite{ENS1}
and watching the condensate not from above, but from one side would allow a
direct observation of bent vortex lines lead to pictures as those in Fig.
(e-h).

\subsection{What is the ultimate cause of bending?}

The fact that the bending of the vortex line takes place not only on the
asymmetric trap of current experiments, but also on an radially symmetric trap,
$\varepsilon=0$ represents a surprising novel type of symmetry breaking in
which the superfluid not only chooses the sense of rotation, but also a plane
for its bending.

It is important to emphasize the counterintuitive nature of the vortex bending
in the case of an isolated vortex line. First, opposite to the case of $^4$He,
the BEC is not constrained by any recipient and there are no asymmetric
boundary conditions which could easily explain the deformation of the
fundamental solutions.  Second, although qualitative reasons for such bending
may be found {\em a--posteriori}, they have never been reported before.  And
even though physical arguments may justify the bending of the vortex line, they
can hardly support the fact that the bent vortex line is a stable stationary
configuration. Both the stability and the stationarity can only be demonstrated
with a direct study of the energy functional.

\begin{figure}
  \begin{center}
    \epsfig{file=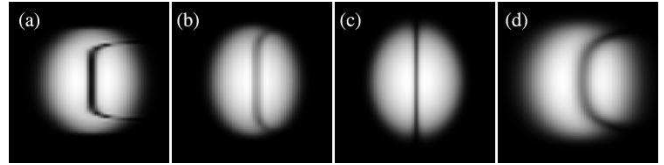,width=\linewidth}
  \end{center}
  \caption{\label{fig-x}
    Side views of the condensate for $N=10^5$ atoms of $^{87}$Rb, and
    decreasing elongation. The trap parameters and plot dimensions
    $(\Omega,\omega_{\perp}/2\pi,\omega_z/2\pi,r_{max},z_{max})$ are
    (a) (0.5$\omega_{\perp}$, 219 Hz, 11.7 Hz, 2.63 nm, 49.09 nm), (b)
    (0.4$\omega_{\perp}$, 96 Hz, 25 Hz, 5.00 nm, 19.23 nm), and (c)
    (0.3$\omega_{\perp}$, 50 Hz, 50 Hz, 8.51 nm, 8.51 nm).  (d) Uses
    the same trap as in (a), but with
    $\{N=10^4,\Omega=0.53,r_{max}=1.65,z_{max}=30.98\}$.}
\end{figure}

Nevertheless, it is legitimate to ask for the ultimate reasons of this rupture
of symmetry. Since our search of states was based on energetical
considerations, we have studied the dependence of the bending with respect to
each of the free parameters in our problem: $N$, $\Omega$, $g$, $\varepsilon$,
and the elongation $\gamma=\omega_\perp/\omega_z$.

First, the influence of $\varepsilon$ is discarded. As we mentioned above,
bending exists in either symmetric or asymmetric traps. Rather, the asymmetry
seems to take part only in the dynamics of the nucleation process, by changing
the value of the critical frequencies and inducing a type of hysteresis
\cite{Ours}.

A different role is played by the elongation of the trap, $\gamma$, and the
effective interaction $U=Ng$. Starting with a configuration and lowering either
the elongation or the number of atoms we see that the bending becomes smaller
and eventually disappears [Fig. \ref{fig-x}(a-c)].

We have tried to measure the bending so as to determine the minimal enlogation
and interaction which are required to induce this phenomenon. The problem is
that due to the small changes of $L_z$, the full minimization procedure does
not allow us to bound this minimal elongation accurately. An alternative
procedure is to study the relation between the angular speed which is required
to stabilize a straight vortex, $\bar\Omega_1$ and the angular speed at which
the ground state acquires some angular momentum, $\Omega_1$. The first value
arises from the study of normal modes around a straight vortex, while the
second value is the energy difference between a straight vortex and a
vortexless state \cite{Ours}. When $\bar\Omega_1\gg\Omega_1$ there are values
of the angular speed where bending can be favorable. This phenomenon has been
referred to in the literature as ``anomalous modes'' \cite{anomalous}, and it
is a signature of bending.

\begin{figure}[htbp]
  \begin{center}
    \epsfig{file=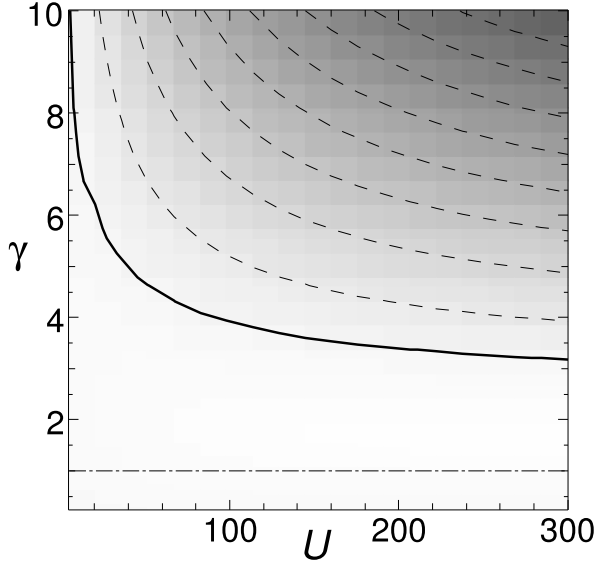,width=0.9\linewidth}
  \end{center}
  \caption{\label{fig-modes}
    Grayscale plot and contour lines for the difference $\bar\Omega_1
    -\Omega_1$ between the speed required to nucleate a vortex, $\Omega_1$, and
    the speed required to stabilize a straight one, $\bar\Omega_1$, as a
    function of the elongation of the trap, $\gamma=\omega_\perp/\omega_z$, and
    the effective interaction, $U=Ng$.  A horizontal dash-dot line marks the
    spherically symmetric trap, $\gamma=1$. For $\bar\Omega_1 > \Omega_1$, i.e.
    above the thick solid line, the bending of vortices becomes energetically
    favorable. All figures are adimensional.}
\end{figure}

In Fig. \ref{fig-modes} we plot the difference $\bar\Omega_1-\Omega_1$, as a
function of the elongation, $\gamma$, and of the effective interaction, $U=Ng$
for an axially symmetric trap. According to this study, a large elongation of
the trap is required to make a bent vortex energetically favorable over a
straight one.

\section{Conclusions}
\label{sec:final}

In this work we have studied vortex lines in a very elongated Bose--Einstein
condensate in a rotating trap. By developing both a customized energy
functional (Sect. \ref{sec:free-energy}) and an optimized descent method (Sect.
\ref{sec:sobolev}) we have been able to find the ground state of a condensate
in a family of different experimental configurations.

The first part of our study (Sect. \ref{sec:bending}) focused on realistic
values of the experimental parameters, taken from the work by Madison et al.
\cite{ENS1,ENS2,ENS3}. The main conclusion of this work is that by increasing
the rotation speed, the condensate achieves states with one, two, three and
more bent vortices. Such vortices form a regular Abrikosov lattice at half the
height of the condensate and deform not very far from the condensate core.

Appart from the observation reported in \cite{abo-shaer} and while current
experiments at the ENS \cite{ENS1,ENS2,ENS3} are carried on a regime in which
vortex lines should bend, this has not been yet observed there. The reason is
that the imaging of a condensate which is a few $\mu$m large requires a
previous expansion in which the cloud becomes disk shaped and bending cannot be
appreciated.  Furthermore, the study of the elongated cloud without vortices
\cite{Ours,Stringari} shows that the nucleation of vortices is subject to
hysteresis and one must actually exceed some rotation frequency $\Omega_m$,
which is rather large and makes it difficult to selectively produce one, two or
more vortices.

On the other hand the bending of vortex lines has signatures which are also
observed in current experiments. First, the bending by itself can explain the
apparent filling of vortex lines when seen from above \cite{ENS1}, and the
regular pictures which appear in most recent observations with sliced sodium
condensates \cite{abo-shaer}. Second the bending also accounts for the
continuous growth of the angular momentum with respect to the rotation speed,
and the fractional values of the angular momentum, $0<l<1$, which arise in the
indirect measures of \cite{ENS2} and \cite{ENS3}. And finally, the nucleation
of many bent vortices and their subsequent interaction during the expansion
phase may lead to turbulent structures that should explain the lack of regular
pictures above a certain angular speed \cite{ENS1,ENS2}.

The second part of this paper studied the dependence of the bending
with respect to the parameters of the mean field model
(\ref{GPE-rot}). Here we conclude that it is both the elongation and
interactions which induce the bending of the vortex line, while the
transverse asymmetry plays no important role.

Our results imply that past studies devoted to the quasilinear limit ($U \ll
1$) would become of no applicability for elongated traps due to the lack of
bending in the simplified models. Second, along this line it would be nice to
find {\em analytically} the geometry of the vortex line as a function of the
interaction $U$ and the elongation $\gamma$.

Our study also reveals that current experiments are being developed in a regime
which is qualitatively different from the setups that have been studied up to
date \cite{Butts,others}.  Values of the relevant parameters $(\gamma,U)$ in
current experiments are so far from most studies that new methods must be
developed to accurately describe these amazing systems.

As an example, let us pose one of the technically difficult questions that
arise in this work. From our numerical work it seems that the extremes of some
of the vortex lines actually reconnect not too far away from the core of the
cloud, forming what is called a vortex ring. Although it is not possible with
current numerical methods to fully support this conjecture, vortex rings have
already been observed in BEC \cite{Ring}. The limits of zero and infinite
radius of a vortex ring lead to a dimensionless zero and an isolated vortex
line, respectively.  Therefore vortex rings would be a nice tool for explaining
the nucleation of vortices and they would allow to interpret the states in this
work as the result of an incomplete nucleation. The proof or refutation of this
conjecture remains a mathematically challenging problem.

This work has been supported by grant BFM2000-0521.


\begin{thebibliography}{99}

\bibitem[\dag ]{emiliojjgr}  {Electronic address: jjgarcia@ind-cr.uclm.es}

\bibitem[\S ]{emiliovmpg}  {Electronic address: vperez@ind-cr.uclm.es}
  
\bibitem{neumann}{Around 1900, John Von Neumann noticed that most mathematical
    models of the date did not take viscosity into account and thus could not
    explain the features of real fluids. He coined the term ``dry water'' to
    despectively refer to those idealized models which did not care for
    dissipation \cite{feynmann}. Bose-Einstein condensates represent an
    experimental realization of such a ``dry fluid'' or superfluid.}
  
\bibitem{nota}{When the superfluid coexists with a normal component, as is the
    case of a condensate for $T>0$, the normal component may rest within the
    vortex.}

\bibitem{BEC}{M. H. Anderson {\em et al}, Science {\bf 269}, 198--201
    (1995); K. B. Davis {\em et al}, Phys. Rev. Lett.{\bf 75}
    3969-3973 (1995).}

\bibitem{JILA}{M. R. Matthews, B. P. Anderson, P. C. Haljan, C. E. Wiemann, E.
    A. Cornell, Phys. Rev. Lett. {\bf 83} 2498-2501 (1999) .}

\bibitem{ENS1}{ K. W. Madison , F. Chevy, W. Wohlleben, J. Dalibard, Phys.
    Rev.  Lett.  {\bf 84} 806-809 (2000).}

\bibitem{ENS2}{F. Chevy, K. W. Madison, J. Dalibard, Phys. Rev.  Lett. {\bf
      85}, 2223-2226 (2000).}

\bibitem{Butts}{D.~A. Butts, D. S. Rokhsar, {\em Nature} {\bf 397}, 327
    (1999).}

\bibitem{others}{T. Isoshima and K. Machida, Phys. Rev. A {\bf 59}, 2203
    (1999); J.~J. Garc{\'\i}a--Ripoll and V.~M. P\'erez--Garc{\'\i}a, Phys.
    Rev. A {\bf 60} 4864 (1999); D. L. Feder {\em et al.}, Phys. Rev. A {\bf
      61} 011601 (2000).}

\bibitem{Fetter}{A.~A. Svidzinsky, A.~L. Fetter, Phys. Rev. A {\bf 62}, 063617
    (2000).}

\bibitem{Ours}{J.~J. Garc{\'\i}a--Ripoll and V.~M. P\'erez--Garc{\'\i}a, Phys.
    Rev. A {\bf 63}, 041603 (2001).}

\bibitem{asym}{J.~J. Garc{\'\i}a--Ripoll and V.~M. P\'erez--Garc{\'\i}a,
    Phys. Rev. A, {\bf 64}, 013602 (2001)}

\bibitem{abo-shaer}{J.~R. Abo--Shaeer, C. Raman, J.~M. Vogels, and W. Ketterle,
    Science (in print).}
  
\bibitem{Sobolev}{J.~J. Garc{\'\i}a--Ripoll and V.~M. P\'erez--Garc{\'\i}a,
    SIAM Jour. Sci. Comput, (in print), e-print
    http://xxx.lanl.gov/abs/math.SC/0008225}
  
\bibitem{nota2}{Such methods do exist, and the so called imaginary time
    evolution is one of them, but it is far from efficient in the most
    realistic problems \cite{Sobolev}.}
  
\bibitem{dipoles}{J.~J. Garc{\'\i}a--Ripoll, V.~M. P\'erez--Garc{\'\i}a, E.~A.
    Ostrovskaya, and Y.~S. Kivshar, Phys. Rev. Lett. {\bf 85}, 82, (2000).}

\bibitem{Neu97}{J.~W. Neuberger, {\em Sobolev Gradients and Differential
    Equations}, Springer-Verlag, Berlin (1997).}

\bibitem{gabriel}{J.~J. Garc{\'\i}a--Ripoll, G. Molina--Terriza, V.~M.
    P\'erez--Garc{\'\i}a, and L. Torner, Phys. Rev. Lett. [submitted
    LD8071]}

\bibitem{anomalous}{D.~L. Feder, A.~A. Svidzinsky, A.~L. Fetter, and
    C.~W. Clark, Phys. Rev. Lett. {\bf 86}, 564 (2001)}

\bibitem{ENS3}{K.~W. Madison, F. Chevy, V. Bretin, and J. Dalibard, e-print
    cond-mat/0101051}

\bibitem{Stringari}{F. Dalfovo, S. Stringari, Phys. Rev. A {\bf 63}, 011601
    (2001).}

\bibitem{Ring}{B.~P. Anderson {\em et al}, e-print cond-mat/0012444.}

\bibitem{feynmann}{R.~P. Feynmann, R. B. Leighton, and M. Sands, ``F{\'\i}sica.
    Vol. II: Electromagnetismo y materia'', Chap. 40, Addison-Wesley
    Iberoamericana, Wilmington (1987).}

\end{thebibliography}
\end{document}